# Monolithic integration of 940 nm AlGaAs distributed Bragg reflectors on bulk Ge substrates

YUNLONG ZHAO,[1] JIA GUO,[1] MARKUS FEIFEL,[2] HAO-TIEN CHENG,[3] YUN-CHENG YANG,[4] LIMING WANG,[1] LUKAS CHROSTOWSKI,[5] DAVID LACKNER,[2] CHAO-HSIN WU,[3,4] AND GUANGRUI (MAGGIE) XIA[1,*]

[1] Department of Materials Engineering, the University of British Columbia, Vancouver BC, V6T 1Z4 Canada

[2] Fraunhofer Institute for Solar Energy Systems ISE, Heidenhofstraße 2, 79110 Freiburg, Germany

[3] Graduate Institute of Electronics Engineering, National Taiwan University, Taiwan, ROC

[4] Graduate Institute of Photonics and Optoelectronics, National Taiwan University, Taiwan, ROC

[5] Department of Electrical and Computer Engineering, University of British Columbia, Vancouver, BC, V6T 1Z4 Canada

[*]gxia@mail.ubc.ca

**Abstract:** High quality 940 nm $Al_xGa_{1-x}As$ n-type distributed Bragg reflectors (DBRs) were successfully monolithically grown on off-cut Ge (100) substrates. The Ge-DBRs have reflectivity spectra comparable to those grown on conventional bulk GaAs substrates and have smooth morphology, reasonable periodicity and uniformity. These results strongly support VCSEL growth and fabrication on more scalable bulk Ge substrates for large scale production of AlGaAs-based VCSELs.

**References**

1. K. Iga, "Surface-emitting laser-its birth and generation of new optoelectronics field," IEEE J. Sel. Top. Quantum Electron. **6**, 1201-1215 (2000).
2. A. Extance, "Faces light up over VCSEL prospects," SPIE Professional magazine, 09 April (2018).
3. M. Dummer, K. Johnson, S. Rothwell, K. Tatah, and M. Hibbs-Brenner, "The role of VCSELs in 3D sensing and LiDAR," Proc. of SPIE **11692**, 116920C (2021).
4. N. Anyadike, "3D sensing technology booming on new applications," Electropages, 30 January (2019).
5. J. F. Seruin, D. Zhou, G. Xu, A. Miglo, D. Li, T. Chen, B. Guo, and C. Ghosh, "High-efficiency VCSEL arrays for illumination and sensing in consumer applications," Proc. of SPIE **9766**, 97660D (2016).
6. H. Moench, M. Carpaij, P. Gerlach, S. Gronenborn, R. Gudde, J. Hellmig, J. Kolb, and A. van der Lee, "VCSEL-based sensors for distance and velocity," Proc. of SPIE **9766**, 97660A (2016).
7. Yole Dévelopment, "VCSEL – Technology and Market Trends 2021 report," Technical report, 2021.
8. D. Wiedenmann, M. Grabherr, R. Jäger, and R. King, "High volume production of single-mode VCSELs," Proc. of SPIE **6132**, 613202 (2016).
9. C. Bocchi, C. Ferrari, P. Franzosi, A. Bosacchi, and S. Franchi, "Accurate determination of lattice mismatch in the epitaxial AlAs/GaAs system by high-resolution X-ray diffraction," J. Cryst. Growth **132**, 427-434 (1993).
10. M. Bosi, and G. Attolini, "Germanium: Epitaxy and its applications," Prog. Cryst. Growth Charact. Mater. **56**, 174 (2010).

11. B. Depuydt, A. Theuwis, and I. Romandic, "Germanium: From the first application of Czochralski crystal growth to large diameter dislocation-free wafers," Mater. Sci. Semicond. Process. **9**, 437 (2006).
12. A. Johnson, A. Joel, A. Clark, D. Pearce, M. Geen, W. Wang, R. Pelzel, and S. W. Lim, "High performance 940nm VCSELs on large area germanium substrates: the ideal substrate for volume manufacture," Proc. of SPIE **11704**, 1170404 (2021).
13. M. Feifel, D. Lackner, J. Ohlmann, J. Benick, M. Hermle, and F. Dimroth, "Direct Growth of a GaInP/GaAs/Si Triple-Junction Solar Cell with 22.3% AM1.5g Efficiency," Sol. RRL. **3**, 1900313 (2019).

**1. Introduction**

Vertical cavity surface emitting lasers (VCSELs), compared to edge emitting lasers, have the advantages of high power efficiency, low beam divergence, narrow spectrum, higher spectral temperature stability, and low-cost testing and packaging [1]. The demands for VCSELs as near infrared illumination sources in three-dimension (3D) sensing and imaging have increased dramatically in recent years as VCSELs power some of the most popular new features in smartphones and augmented reality (AR) applications, such as Face ID, Animoji and proximity-sensing [2-3]. They are also widely used in auto-navigation, smart retail, industrial automation and robotics. The global market for VCSELs was predicted to increase to $2.4 billion by 2026 from the $1.2 billion market size in 2021 at a 13.6% compound annual growth rate [4-7]. So far, VCSEL production is still largely based on 3-, 4- and 6-inch bulk GaAs wafers [8]. Scaling up the substrate wafer size is a very effective way to boost the production volume, lower the manufacturing cost and make larger VCSEL arrays more ubiquitous.

The most common emission wavelengths of VCSELs are in the range of 750–1100 nm. A typical VCSEL epitaxial structure is about 5 to 15 microns thick, with the majority of the thickness taken by the bottom and the top distributed Bragg reflectors (DBRs) made of $Al_xGa_{1-x}As/Al_yGa_{1-y}As$ superlattices. The lattice constant mismatch is about 0.14% between AlAs and bulk GaAs at room temperature, which is acceptable for VCSEL production based on 3-inch and 4-inch GaAs wafers [9]. When the growth is transited to larger GaAs wafers, the inherent strain formed during VCSEL growth results in bowed and warped wafers, which lead to low chip yield and reliability problems. One solution is to replace the bulk GaAs substrates with bulk Ge substrates, as the lattice constant of Ge locates between those of GaAs and AlAs. The less lattice mismatch between AlAs and Ge allows VCSELs epitaxy with lower residual strain, which reduces bow and warp after growth on Ge wafers [10]. With neglectable bow and warp, thicker DBR epilayers for longer wavelength VCSELs and larger diameter substrates can be used. Ge wafers also have some other attractive properties, including much lower etch pit density (EPD) and dislocations, which are associated with yield loss and reliability failures in VCSELs on bulk GaAs substrates [11]. The price of mass-produced 6" Ge wafers is not higher than 6" GaAs wafers. Most importantly, Ge wafers can be made to 8- and 12-inch diameter, which is very hard to achieve for more brittle GaAs wafers. All these arguments make Ge wafers promising to replace GaAs wafers as the substrates for the mass production of VCSELs. In 2020, IQE first demonstrated 6" bulk Ge wafer based 940 nm VCSELs [12]. With unoptimized processes for Ge-VCSELs, they produced comparable DBR stopbands, active region photoluminescence (PL), lasing performance as GaAs-VCSELs and have much less wafer bow and slip [12]. It is the only report on bulk Ge-based VCSELs so far. However, no Ge substrate specifications, Ge to GaAs transition layers, DBR growth and testing conditions were revealed in their presentation.

In this paper, we report the successful monolithic integration of $Al_xGa_{1-x}As$ n-type distributed Bragg reflectors (n-DBRs) on bulk Ge substrates, which is the first and crucial step of bulk Ge-based VCSELs. Important material, processing and testing conditions are discussed in detail.

2. **Epitaxy growth**

The schematic structure of the DBRs grown on Ge substrates is shown in Fig. 1. The Ge wafers used in this study were n-type 375 μm thick 4-inch (100) Ge wafers with 6-degree miscut towards (111) provided by Umicore. This thickness was chosen to match the 3" Si adaptor thickness so that the surfaces of the Ge samples are flush with the Si adaptor surface. Due to the more mechanical robustness, Ge wafers can be thinner than GaAs wafers of the same sizes. The miscut is to reduce antiphase-domains of the subsequent GaAs layers on Ge and to promote step-flow growth. 100 nm SiN thin films were deposited on the Ge wafer backsides to prevent Ge evaporation.

The AlGaAs n-DBR superlattices are not directly on top of the Ge surface. Instead, an InGaP nucleation layer, a 950 nm n-$Ga_{0.985}In_{0.015}$As layer lattice matched to Ge and a 50 nm n-GaAs layer with $5\times10^{18}$ cm$^{-3}$ doping were grown as a nucleation layer on Ge in order to have a clean and anti-phase free surface. The growth was performed in an Aixtron 2800G4R metal organic chemical vapor deposition (MOCVD) reactor and the details are in [13]. After the backside SiN deposition and frontside $Ga_{0.985}In_{0.015}$As/GaAs layer growth, the 4" Ge wafers were 376 μm in thickness. They were cut into 1 cm × 1 cm pieces. 3-inch 380 μm thick Si dummy wafers with 1 cm × 1 cm square holes cut by laser machining were used to position the Ge pieces in the 3" wafer pockets of a MOCVD reactor at LandMark Optoelectronics Corp for the n-DBR growth. First, a 500 nm thick GaAs layer was grown to have a cleaner surface for the subsequent n-DBRs. A 3" 2-degree offcut bulk 629 μm thick GaAs wafer was used as the substrate of the control sample. Each DBR consists of 40 periods of n-type $3\times10^{18}$ cm$^{-3}$ doped 20 nm Al$_x$Ga$_{1-x}$As (x: 0.12→0.9) / 56 nm Al$_{0.9}$Ga$_{0.1}$As / 20 nm Al$_x$Ga$_{1-x}$As (x: 0.9→0.12) / 49 nm Al$_{0.12}$Ga$_{0.88}$As, which was designed with 940 nm as the center wavelength of the reflectance band. The growth temperature was set to 760 °C and the vapor pressure was 50 mbar. For the DBR growth process, the sources used for GaAs growth was H$_2$, SiH$_4$, AsH$_3$ and Ga(CH$_3$)$_3$ (TMGa), which is the preferred metalorganic source of gallium for MOCVD. During the Al$_x$Ga$_{1-x}$As growth, Al$_2$(CH$_3$)$_6$ (TMAl) was added to provide the Al source.

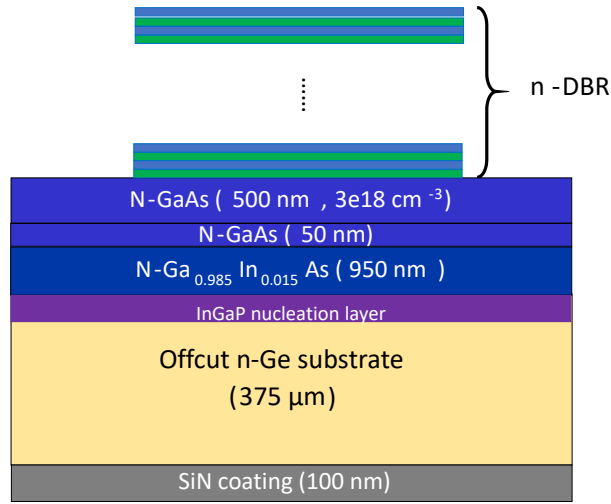

Fig. 1. Schematic structure of the DBRs grown on Ge substrate with GaAs buffer layer.

## 3. Characterization results and discussions

Good optical and material properties of the DBRs, i.e. reflectance spectra with stop band, smooth morphology and low threading dislocation density, are essential for the operation of VCSELs. To investigate the impact of the Ge substrates on the DBR quality, atomic force microscope (AFM) imaging, optical reflectance spectra measurements, high-resolution X-ray diffraction (HRXRD), scanning electron microscopy (SEM), etch pit density (EPD), and electron channeling contrast imaging (ECCI) analysis were used.

### 3.1 Surface quality

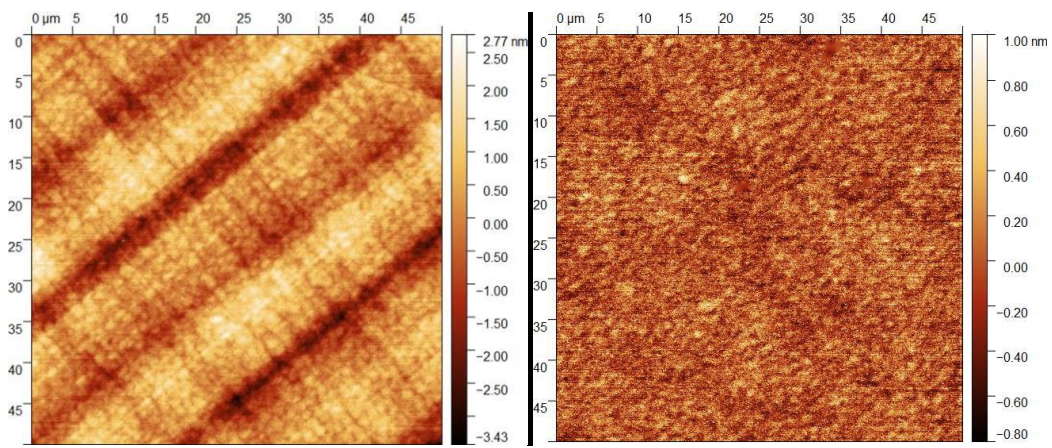

Fig. 2. AFM images of the DBR surfaces on (a) a Ge substrate and (b) on a GaAs substrate.

Under the optical microscope and AFM, both Ge-DBRs and GaAs-DBRs show good surface quality, which are clean and free of cracks and antiphase domains. Some minor cross-hatch pattern can be seen in the AFM image of the Ge-DBR surface, as shown in Fig. 2, which is a result of some residual strain in the epitaxial layers due to the small lattice mismatch between AlGaAs, GaAs and Ge. The root mean square (RMS) roughness measured by AFM for Ge-DBRs and GaAs-DBRs are 0.83 and 0.28 nm respectively measured on 50 µm x 50 µm areas. The RMS roughness of the GaAs/InGaAs/InGaP/bulk-Ge/SiN substrates is 0.43 nm before the DBR growth.

**3.2 Reflectance**

Fig. 3 shows the normal-incidence reflectance spectra of the Ge-DBR and GaAs-DBR, which were obtained with a Filmmetrix F20 thin-film analyzer. Reflectance measurements are sensitive to the position and tilt of the surface. A rotation stage was used underneath the samples to check the reflectance variation brought by the small surface tilt. Each DBR was measured with a series of rotation angles from 0 to 180 degrees. The highest reflectance was recorded, when the sample surface, the incident light and the reflection light detector satisify the law of reflection. Ge-DBR has a stronger rotation angle dependence with 5.28% variation due to the rougher surface and larger non-uniformity, compared with 4.17% variation for bulk GaAs-DBR.

Fig. 3a shows that Ge-DBRs have comparable stop-band shapes, widths and maximum peak heights as those of the control GaAs-DBRs. The as-collected peak reflectance values of the Ge-DBRs and the GaAs-DBRs are in the range of 113.31 to 115.45 and 111.03 to 114.94 respectively. These reflectance values are normalized to a Si calibration sample of Filmmetrix. The peak reflectance values of the Ge-DBRs are from 98.58 % to 100.44 % of the peak GaAs-DBR reflectance. The errors mainly come from the non-uniformity of the DBR growth and the measurement method.

The stop-band width of the Ge-DBRs and the bulk GaAs-DBRs are 74.4 nm and 77.3 nm respectively. This difference is within the cross-wafer deviations of the Ge-DBRs shown in Fig. 3(b). There is a noticeable spectral shift between the Ge-based and GaAs-based samples. The stop-band centers of GaAs-DBRs locate at 940 nm, while those of the Ge-DBRs range from 920 nm to 940 nm.

SEM microscope images taken at different locations reveal that the average Ge-DBRs and GaAs-DBRs are 5848 nm and 5924 nm respectively (see section 3.4), which means that the Ge-DBR layers are about 1.45% thinner than those of the GaAs-DBRs. This is consistent with the smaller average center wavelength of the Ge-DBR stopband. Although the 629 µm thick GaAs wafer and the 376 µm thick Ge pieces were processed in the same DBR epitaxy growth, the Ge pieces were fixed by Si dummy adaptors, while the GaAs substrate stayed in the wafer pocket as a full 3" wafer. The difference in heat conduction and gas flow could result in different growth rate and uniformity of the Ge-DBRs compared with the GaAs-DBRs.

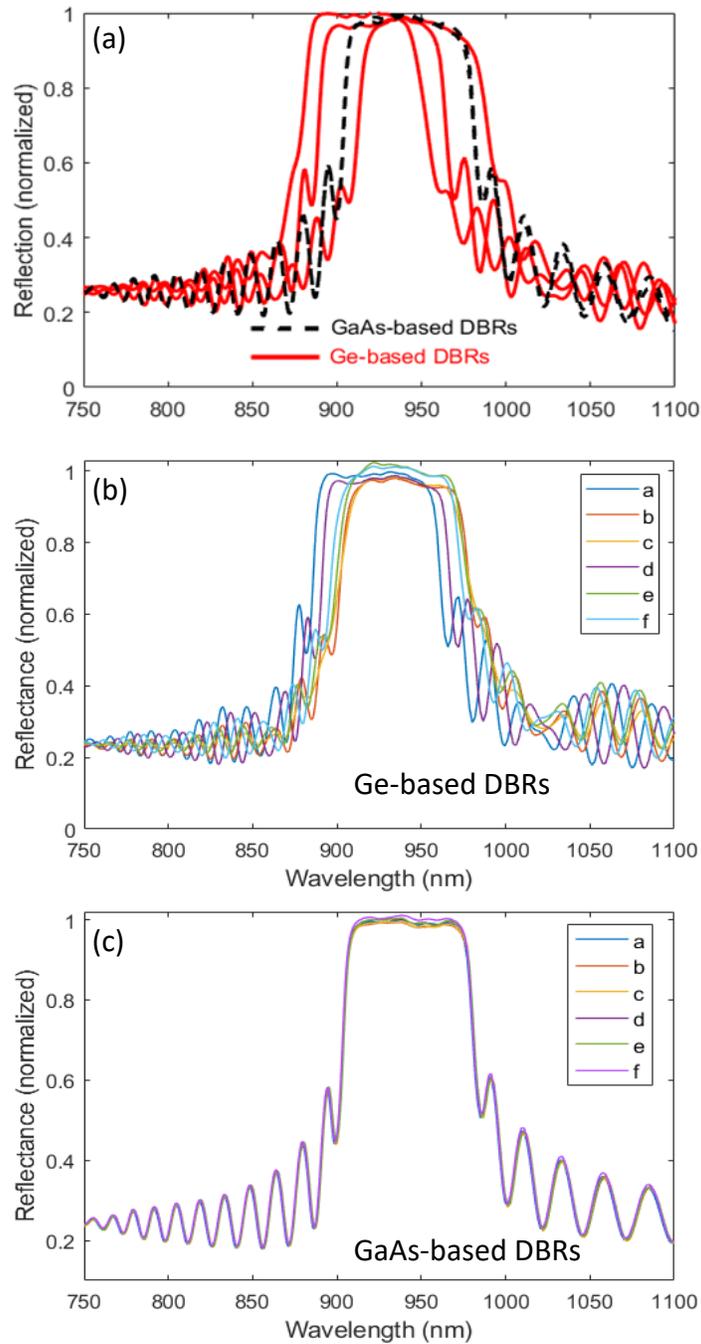

Fig. 3. (a) Normal-incidence reflectance spectra of bulk GaAs-based DBRs and Ge-based DBRs grown at the center of 3-inch Si dummy adaptor. Both spectra are taken at the center of each sample normalized by the maximum GaAs DBR reflectance value. (b) Normal-incidence reflectance spectra of 6 pieces of Ge-based DBRs (a to f) placed at different positions of 3-inch Si dummy adaptor showing the cross-wafer growth non-uniformity, which are normalized by the maximum Ge-DBR reflectance of the spectrum from the center spot growth "a" (the blue line). (c) Normal-incidence reflectance spectra of bulk GaAs-based DBRs measured at 6 spots (a to f) across a 3-inch bulk GaAs wafer showing the cross-wafer reflectance uniformity, which are normalized by the maximum GaAs DBR reflectance of the spectrum from the center spot "a" (the blue line).

### 3.3 X-ray diffraction

In order to check the quality of the epitaxial growth of Ge-DBRs, HRXRD ω/2θ rocking curves were measured for both Ge-DBRs and control GaAs-DBRs. Ge (220) 4-bounce monochromator for incident beams and a 1 mm slit for the detector were used to characterize both structures.

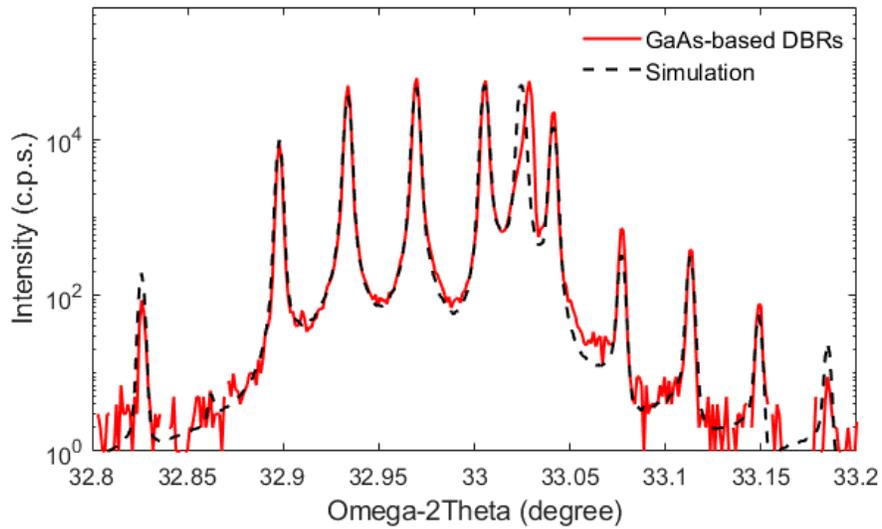

Fig. 4. (004) omega-2theta scan of bulk GaAs-DBRs in comparison with the simulated structure.

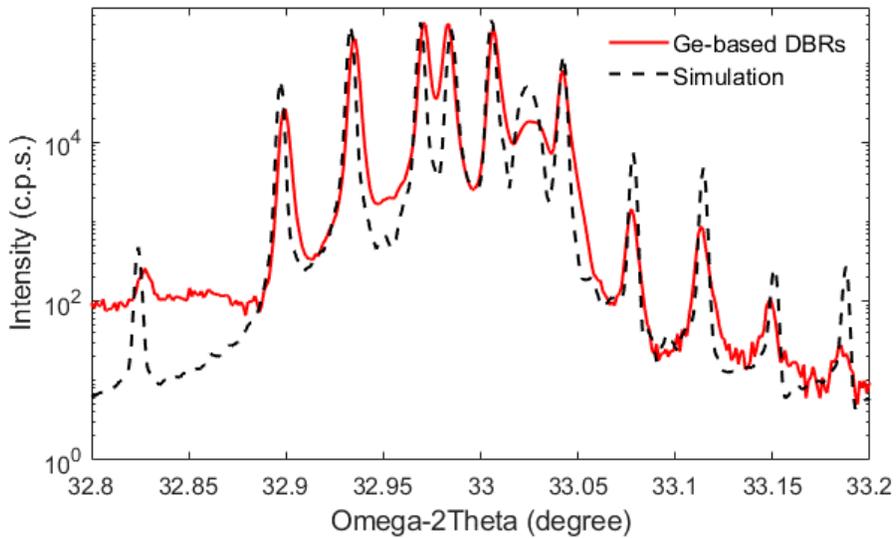

Fig. 5. (004) omega-2theta scan of Ge-DBRs in comparison with the simulated structure.

In Fig. 4 and 5, the (004) ω/2θ diffraction curves of the control GaAs-DBR and Ge-DBR are plotted in comparison with the simulated curves respectively. The simulation was conducted by X'Pert Epitaxy 4.1 software. The best-fitting GaAs-DBR rocking curve in Fig. 4 was based on a uniform DBRs structure with 5.93 µm thickness, and the best-fitting Ge-DBR rocking curve in Fig. 5 was based on a uniform DBR structure with 5.84 µm thickness. The difference in the best-fitting thickness is consistent with the measured results.

In Fig. 4, a very good agreement on the peak positions and intensity between the simulated and measured GaAs-DBR can be observed, indicating very good DBRs thickness uniformity and crystal quality. The slight shift of GaAs peak from 33.0239 degree in simulation to 33.0287 degree in measurement is from the strain in the grown buffer GaAs layer beneath the DBRs, which is not included in the simulation. In comparison, in Fig. 5, the measured peaks from the Ge-DBR are slightly broader leading to minima that are less defined. The peaks and satellite-peaks occur still at the same positions, which means that the material compositions are as designed and a good epitaxy growth quality is achieved. The broader and lower peaks, however, indicate that there is more variation in the layer thickness, surface roughness due to the strain relaxation (as seen in the AFM images in Fig. 2), and a higher density of crystal defects like threading dislocations, which were confirmed by SEM imaging and dislocation density measurement discussed below.

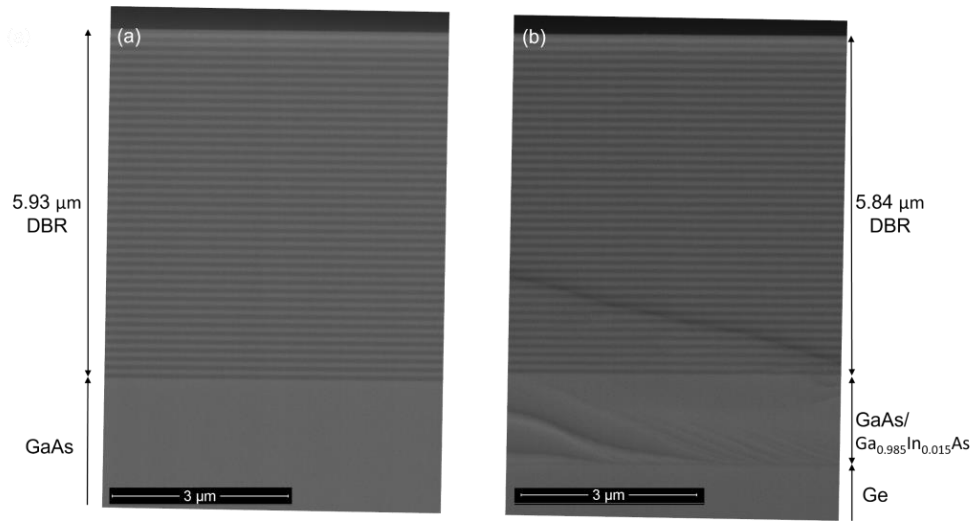

Fig. 6. Cross-section SEM images at 30K × magnifications of a (a) bulk GaAs-based DBR, (b) Ge-based DBR.

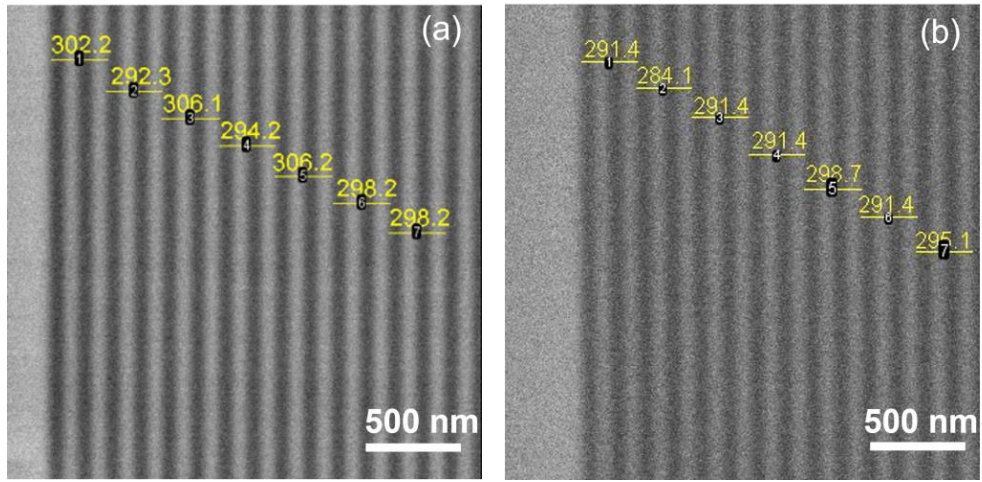

Fig. 7. Epitaxial DBR layer thickness distribution of (a) bulk GaAs-based DBR, (b) Ge-based DBR.

**3.4 Cross-sectional SEM images**

SEM imaging of the cross-sections of the Ge-DBRs and GaAs-DBRs was conducted to observe the DBR layers. The SEM samples were manually cleaved with a diamond scriber. No surface polishing or coating was conducted. The images in Fig. 6 and 7 were collected using a FEI Nova NanoSEM system. The measurement mode was the immersion mode with a gaseous analytical detector (GAD). The operation voltages used were 10 to 15 kV and the working distance was 5–6 mm. SEM images of the Ge- and GaAs-DBR surfaces (not shown) reveal smooth surfaces without any cracks or surface defects within the imaging area of about $7.5 \times 10^5$ μm$^2$. The SEM images of the cross-sections of bulk GaAs- and Ge-DBRs in Fig. 6(a) and Fig. 6(b) show good periodicity and uniformity for both types of DBRs. The shadowy lines in Fig. 6(b) are due to the roughness from the sample cleaving. A noticeable difference between the two images is on the thickness of the DBR. The thickness of the bulk GaAs-DBR is 5.93 µm, while that of the Ge-DBRs is 5.84 µm. It is 1.5% thinner, consistent with the shorter center wavelengths of the Ge-DBRs. The thicknesses of every two repeated pair of a GaAs-DBR and a Ge-DBR are shown in Fig. 7(a) and Fig. 7(b), and that of the Ge-DBR is 1 to 5% thinner than that of the GaAs-DBR. This thickness difference was not unexpected, as the DBR growth conditions used were optimized for GaAs-DBRs to obtain optimal yield, uniformity and VCSEL performance at LandMark. Ge substrates are 376 µm thick in comparison with 629 µm thickness of the GaAs substrates, which may have led to small variations in thermal coupling and finally slightly different surface temperatures. Also, Ge samples are much smaller (1 cm by 1 cm), and were placed inside the Si adaptors. Due to the edge effects, the non-uniformity of the smaller Ge-DBRs is much larger than that of the full GaAs-DBRs. To avoid these problems, in the next epitaxy growth, 3" full Ge wafers will be used. Moreover, the MOCVD growing conditions will be fine-tuned for Ge-DBRs to further improve the thickness, uniformity and stop-band characteristics.

**3.5 Dislocation density measurement**

Threading dislocations (TDs) are key defects that degrade laser efficiency and lifetime. EPD and ECCI measurements were conducted on the Ge substrates before and after the DBR growth respectively to measure the threading dislocation density (TDD). The two methods were chosen according to their suitable TTD ranges. The EPD etchant solution consisted of $CH_3COOH$ (100 ml), 70% by weight $HNO_3$ (40 ml), 49% by weight HF (10 ml), and $I_2$ (30 mg). The ECCI measurement was performed with a SEM (Apero 2 model by ThermoFisher Scientific) using the PivotBeam mode. The crystal orientation was optimized using the sample stage tilt and rotation, and crystal defects imaging was conducted in the Immersion mode. All ECCI imaging was performed in high vacuum and no conductive coating was applied. The GaAs/InGaAs/InGaP/bulk-Ge/SiN substrate (before the DBR growth) has a TDD of about $9.6 \times 10^2$/cm$^2$ obtained over a total area of 13.3 mm$^2$ by EPD and the Ge-DBR has a TDD of about $2.1 \times 10^6$/cm$^2$ obtained over a total area of 144.6 μm$^2$ by ECCI.

Our next growths of partial or full VCSEL epitaxy on Ge wafers have been planned. Calibration runs will be performed to tune the Ge-DBRs growth recipe to have the stopbands centered at 940 nm. Full Ge wafers are going to be used to avoid growth condition variation and improve the growth uniformity and quality. More material analysis and device performance measurements will be conducted.

4. Conclusion

High quality AlGaAs DBRs were successfully monolithically grown on bulk Ge substrates. The Ge-DBRs have reflectivity spectra comparable to those grown on conventional bulk GaAs substrates and have smooth morphology, and reasonable periodicity and uniformity. No APDs are formed in DBR epitaxy on Ge substrates, and the threading dislocation density of Ge-DBR is about $2.1 \times 10^6$/cm$^2$. These results are relevant to the R&D of VCSEL growth and fabrication on large-area Ge substrates to address the dramatically increased demand on VCSELs.

5. Acknowledgment

Huawei Technologies, Canada is acknowledged for funding this project. Umicore N. V., Belgium, is acknowledged for providing the bulk Ge wafers with GaAs/GaPAs epitaxy layers and SiN back coating. ThermoFisher Scientific corporation was acknowledged for the ECCI measurements.